\documentclass[12pt]{article}

\usepackage{graphicx}
\usepackage{times}
\usepackage{natbib}
\usepackage{fancyhdr}
\usepackage{color}
\usepackage{amsmath}
\usepackage{epsfig}
\usepackage{multirow}

\renewcommand{\baselinestretch}{1.5}
\begin{document}

\bibliographystyle{nature}
%


\renewcommand{\multirowsetup}{\centering}

\title{Laplacian Dynamics and Multiscale Modular Structure in Networks}

\author{R. Lambiotte$^{1}$,  J.-C. Delvenne$^{2,3}$ and M. Barahona$^{1}$}
\maketitle

\begin{center}
\small{$^1$ Institute for Mathematical Sciences, Imperial College London, 53 Prince's Gate, London SW7 2PG, United Kingdom\\
$^2$Department of Mathematical Engineering, Universit{\' e} catholique de Louvain, 4 avenue Georges Lema{\^i}tre, B-1348 Louvain-la-Neuve, Belgium\\
$^3$ D{\' e}partement de Math{\' e}matique, Facult{\' e}s Universitaires Notre-Dame de la Paix, 8 rempart de la vierge, B-5000 Namur, Belgium}
\end{center}

\noindent {\bf
Most methods proposed to uncover communities in complex networks rely on
their structural properties. Here we introduce the stability of a network
partition, a measure of its quality defined in terms of the statistical properties of a dynamical process taking place on the graph. The time-scale of the process acts as an intrinsic parameter that uncovers community structures at different resolutions. The stability extends and unifies standard notions for community detection: modularity and spectral partitioning can be seen as limiting cases of our dynamic measure. Similarly, recently proposed multi-resolution methods correspond to linearisations of the stability at short times. The connection between community detection and Laplacian dynamics enables us to establish dynamically motivated stability measures linked to distinct null models. We apply our method to find multi-scale partitions for different networks and show that the stability can be computed efficiently for large networks with extended versions of current algorithms.
}

\maketitle

The relation between the structure of a network and the
dynamics that takes place on it has been studied extensively in the last
years~\cite{review,newman_review,syn2}.  A growing body of research has shown how processes such as diffusion and synchronisation are affected by the underlying
graph topology and, conversely, how such dynamical processes can be used
to extract information about the network and reveal its structural properties. This two-way relationship is particularly relevant
when the network is composed of tightly-knit modules or \textit{communities}~\cite{GN,simon,fort,revporter,
sales,clauset}.
On the one hand, it has been observed numerically that the presence of communities enhances partially coherent dynamics in networks~\cite{syn1},
and has a direct effect on the emergence of co-operation~\cite{cooperation} and the coexistence of heterogeneous ideas in a social network~\cite{lambi}.
On the other hand, dynamical processes such as random walks~\cite{PL,rosvall} and synchronization~\cite{syn1} have been proposed as empirical means to uncover the community structure of networks.

In parallel to these dynamical studies, there has been extensive research on community detection based on \textit{ structural} properties of graphs~\cite{fort,revporter}. Several methods have been proposed in order to partition a network into communities, as a way to coarse-grain the level of description of the system but also to identify underlying, often unknown, functionalities or relationships between the nodes \cite{amaral_nat}. At the heart of these methods, there is always a definition for what is thought to be the goodness of a partition. In the last decades, a variety of quality functions have been proposed, such as cut, normalized cut, modularity and extensions of modularity, together with heuristics in order find the partition optimizing the particular quality function. These definitions have in common to be combinatorial, in the sense that the quality of a partition is always measured by counting the number of links within/between the communities. A good example is the widely-used modularity\cite{newman_modul_PNAS,NG} of a partition $\mathcal P$, which 
measures if there are more edges within communities than would be expected on the basis of chance.
In the case of undirected and unweighted networks, modularity reads
\begin{equation}
\label{modularity}
Q = {1\over2m} \sum_{C \in \mathcal{P}} \sum_{i,j \in C} \biggl[ A_{ij} - P_{ij} \biggr]
\quad \quad
\begin{cases}
\text{if $P_{ij}=\langle k\rangle^2/2$,} & \text{then $Q \equiv   Q_{\rm unif}$} \\
\text{if $P_{ij}= k_i k_j/2m$,} &  \text{ then $Q \equiv  Q_{\rm conf}$.}
\end{cases}
\end{equation}
Here, $A$ is the adjacency matrix, where $A_{ij}$ is equal to 1 if there is a link between node $i$ and node $j$, and zero otherwise;
the degree of node $i$ is defined as $k_i \equiv \sum_{j} A_{ij}$; and
$m \equiv \sum_{i,j} A_{ij}/2$ is the total number of links in the network. The summation is performed over all pairs of nodes belonging to the same community $C$ of the partition $\mathcal P$. Effectively, modularity therefore counts the number of links within communities and compares it to the expected number of such links in an equivalent null model. The standard choices for $P_{ij}$ are the uniform model $P_{ij}=\langle k\rangle^2/2$ and the Newman-Girvan model $P_{ij}= k_i k_j/2m$, which is closely related to the configuration model and is often preferred because it takes into account the degree heterogeneity of the network~\cite{lastn}. Despite some limitations, modularity $Q$ has revealed a crucial quantity which is now at the heart of several community detection methods, and has been generalised to weighted and/or directed networks (see Supplementary Information). 

\section*{Stability of a partition}

The purpose of this paper is to clarify the relationship between dynamics and structure by proposing a general framework in order to measure the quality of a partition in terms of the properties of a stochastic process applied on the graph. To do so, we generalize a recent approach where the quality of the partition of a graph is expressed in terms of its stability, an autocovariance function of a Markov process on the network~\cite{JC}. Without loss of generality, let us describe some Markov process $\mathcal M$ in terms of the motion of a random walker moving on the graph. Under the single condition that $\mathcal M$ is ergodic, namely that any initial configuration will asymptotically reach the same stationary solution, the stability of the partition $\mathcal{P}$ ~ is defined as
\begin{equation}
R_{\mathcal M}(t) =\sum_{C \in \mathcal{P}}  P(C,t) - P(C,\infty)
\label{eq:modDeff}
\end{equation}
where $P(C,t)$ is the probability for a walker to be in the same community $C$ initially and at times $t$ when the system is at stationarity. Because the dynamics is ergodic, the memory of the initial condition is lost at infinity, and $P(C,\infty)$ is therefore also the probability for two independent walkers to be in $C$. Stability measures the quality of a partition in terms of the persistence of the dynamics by giving a positive contribution to communities from which a random walker is unlikely to escape within the given time scale $t$.

Stability differs from modularity in several aspects that we will develop throughout this paper. First, stability is based on flows of probability on the graph and therefore captures how the global structure of the system constrain patterns of flows, while modularity focuses on pairwise interactions and is blind to such patterns, thereby neglecting important aspects of the network architecture \cite{rosvall}.
Second, stability describes the quality of the partition of a graph at different time scales and this quantity is, in general, optimised by different partitions at different times. The measure $R_{\mathcal M}(t)$ can thus be used as an objective function to be maximised at each time, thereby leading to a sequence of optimal partitions at different times. We will show below that time is a resolution parameter that allows to unravel the structure of the network at different scales, and that several heuristics measures correspond to the stability of some process at a certain time scale.  
 Finally, our work shows that different Markov processes applied on the same graph lead to different versions of stability, thereby leading to different optimal partitions. This result underlines that the definition for a good partition into communities should depend on the nature of the network and of the dynamics taking place on it.
The flexibility in the definition of $R_{\mathcal M}(t)$ has therefore the advantage to allow a user to chose a particular process, e.g. modelling how information or energy flows on the graph, in order to define an adapted quality function. 

\section*{Undirected vs directed networks}

In order to clarify the above concepts, let us first focus on the discrete-time dynamics of an unbiased random walker, which is a prototype model for diffusion of information. In general, the density of random walkers on node $i$ at step $n$, denoted by $p_{i;n}$, evolves according to:
\begin{equation}
\label{discrete}
p_{i;n+1} = \sum_{j} \frac{A_{ij}}{k_j^{out}} p_{j;n},
\end{equation}
where $A_{ij}$ is now the weight of the link going from $j$ to $i$, and $k_j^{out}=\sum_i A_{ij}$ is the out-strength of node $j$.
The right-hand side accounts for the walkers situated at $j$ at time $n$ and arriving at $i$ by following a link going from $j$ and $i$. In this process, the probability of a walker to move between two nodes is proportional to the weight of the link. 

For a generic undirected network\footnote{We assume the system to be that ergodic, i.e., the network is connected and non-bipartite, in order to avoid unnecessary technical complications.}, for which the adjacency matrix is symmetric $A_{ij}=A_{ji}$ (and therefore $k_i^{out}=k_i^{in}=k_i$),
it is known that the stationary solution of the dynamics is given by $p_i^*=k_i/2m$, where $m$ is now the total weight in the network. For such a process, the probability that a walker is in a community $C$ at stationarity is  $\sum_{j \in C}
k_j/2m $ while the probability of a random walker to be in $C$ during two successive time steps is  $ \sum_{i,j \in C}  (A_{ij}/k_j) (k_j/2m)$.
It follows that stability at time 1 is given by
 \begin{equation}
R_{RW}(1)= \sum_C \sum_{i,j \in C} 
 \left[ \frac{A_{ij}}{k_j} \frac{k_j}{2m} -  \frac{k_i}{2m}  \frac{k_j}{2m} \right] =  Q_{\rm conf} , 
\end{equation}
which is equal to the (configuration) modularity.
Interestingly, the configuration null model 
appears naturally from the properties of the random walk, and not as an additional choice, as in the static definition~(\ref{modularity}). This 
Markov viewpoint therefore provides a dynamical interpretation of modularity, which can be seen as 
the special case (based on paths of length one) of the stability $R_{RW}(t)$ of the partition, a more general quality measure which is now a function of time\cite{JC}.

This connection between modularity and stability is only valid in the case of undirected networks.  To show so, let us now focus on the same random process (\ref{discrete}) applied to a directed network. We further assume that the network is strongly connected\footnote{If this is not the case, it is always possible to make the dynamics ergodic by introducing "\`a la Google" random teleportations\cite{rosvall}.}, in order to ensure that the dynamics is ergodic.
In that case, the stationary density of walkers is given by the (normalized) dominant eigenvector $\pi$ of the transfer matrix $\frac{A_{ij}}{k_j^{out}}$. By repeating the above steps, one finds that stability at $t=1$ is given by\footnote{The quality function (\ref{R1}) has been proposed independently during the revision of this paper by Kim et al.\cite{fKim}}
 \begin{equation}
 \label{R1}
R_{RW}(1)=  \sum_C \sum_{i,j \in C} 
\left[ \frac{A_{ij}}{k_j^{out}} \pi_j -  \pi_i \pi_j \right],
\end{equation}
which is conceptually different from the combinatorial definition (\ref{modularity}), and from its generalization for directed networks. Indeed, (\ref{R1}) favours partitions with long persistent flows of probability within modules, while modularity favours instead partitions with high densities of links (see Supplementary Information). In the case of undirected networks, these two aspects reconcile but they may lead to different and complementary ways to analyse the same graph when the links are directed.

\section*{Time as a resolution parameter}

Further insight into our dynamical interpretation can be gained from
continuous-time processes associated with the random walk~(\ref{discrete}).
If we assume that there are independent, \textit{identical} homogeneous Poisson processes defined on each node on the graph, such that the walkers jump at a constant rate from each node, the corresponding continuous-time process is governed by the Kolmogorov equation
\begin{equation}
\label{ctrw}
\dot{p}_{i} = \sum_{j} \frac{A_{ij}}{k_j} p_{j} - p_i,
\end{equation}
driven by the operator $A_{ij}/k_j - \delta_{ij}$, which is related by similarity to the \textit{ normalised Laplacian} matrix. Here, we have adopted the notation $k_i^{out}=k_i$ as we will only focus on undirected networks from now on.
Note that the stationary solution of this process is also $p_i^*=k_i/2m$.
Following the above discussion, we define the continuous-time stability of
a partition under such dynamics as the probability of a walker to be in the same community after a time $t$ discounting the probability of such an event to take place by chance at stationarity:
\begin{equation}
\label{Rt}
R_{\rm NL}(t) =   \sum_C \sum_{i,j \in C} \biggl[ \left( e^{t (B-I)} \right)_{ij} {k_j\over2m} - {k_i\over2m} {k_j\over2m} \biggr].
\end{equation}
Here $B_{ij}=A_{ij}/k_j$ and $I$ is the identity matrix.
It is possible to show that $R_{\rm NL}(t)$ is a convex, non-increasing function and that it goes to zero for any partition when $t \rightarrow \infty$. This
means that the memory of the initial condition is lost at infinity.  

The role played by time in this measure can be understood by considering the limiting behaviour of $R_{\rm NL}(t)$. Let us first focus on the case $t=0$
\begin{eqnarray}
\label{R0}
R_{\rm NL}(0) =  1 -  \sum_C \sum_{i,j \in C} {k_ik_j\over(2m)^2},
\end{eqnarray}
where it is easy to see 
that $R_{\rm NL}(0)$ is maximised when  each of the nodes is in its own community,
i.e., the optimal partition at $t = 0$ is therefore the finest possible.
On the other hand, in the limit $t \rightarrow \infty$,  it can be shown via eigenvalue decomposition (see Supplementary Information) that $R_{\rm NL}(t)$ is typically maximised by a partition into two communities in accordance with the normalized Fiedler eigenvector, a classic method in spectral clustering~\cite{shimalik}.
The optimisation of the stability $R_{\rm NL}(t)$ therefore leads to a sequence of partitions where the number of communities typically decreases as time grows, from a partition of $N$ one-node communities at $t=0$ to a two-way partition as $t \to \infty$. It is in this sense that time can be seen as an intrinsic resolution parameter in our measure: as time grows, the size of the communities is adjusted to reveal the possible hierarchical structure present in those networks for which finding  just one partition is 
not satisfactory~\cite{simon,clauset,sales,stability2}.

This flexibility is also important to address the so-called ``resolution limit" of modularity~\cite{FB}, namely the fact that modularity optimisation may fail to identify the natural clustering of a network but instead give
too coarse a partition (see Supplementary Information for a longer discussion). 
In our stability framework, partitions beyond the resolution limit are obtained for small time, as we approach $R_{\rm NL}(0)$, where the optimal partition is the finest possible. This $t\to 0$ limit has interesting properties. Firstly, the slope of $R_{\rm NL}(t)$ at the origin equals the cut fraction, i.e., the fraction of edges with extremities in different communities: $-dR_{\rm NL}/dt|_{t=0} = {\rm Cut}
= R_{\rm NL}(0)-  Q_{\rm conf}$. Hence partitions based on MinCut have the slowest decay of $R_{\rm NL}(t)$ at the origin. Secondly, keeping linear terms in $t$ in the short time expansion of $R_{\rm NL}(t)$ leads to
\begin{equation}
\label{Rsmall}
R_{\rm NL}(t) \approx  (1-t)\, R_{\rm NL}(0) + t \, Q_{\rm conf} \equiv Q_{\rm NL}(t).
\end{equation}
The approximate stability $Q_{\rm NL}(t)$ is a convex combination of $R_{\rm NL}(0)$ and the
configuration modularity such that $Q_{\rm NL}(1)=  Q_{\rm conf}$. In fact, $Q_{\rm NL}(t)$ is equivalent up to a linear transformation to the tuneable Hamiltonian proposed by Reichardt and Bornholdt~\cite{reichardt} as an \textit{ad hoc} measure to obtain partitions beyond the resolution limit. Our analysis shows that this measure can be interpreted in terms of a simple linear approximation of the stability $R_{\rm NL}(t)$. However, it is important to remark that the stability $R_{\rm NL}(t)$ provides a more complete picture of the underlying topology of the network at different scales, since it takes into account paths of any length between pairs of nodes. Such exploration of longer paths is essential if one is to detect groups of nodes characterised by longer connectivity patterns~\cite{encore}.

\section*{Other Markov processes}

The stability $R_{\rm NL}(t)$ emerges from the statistical properties of the normalised Laplacian dynamics~(\ref{ctrw}), which can be seen as a generic paradigm for the diffusion of some conserved quantity, e.g., information, on the network. However, many physical processes are instead
governed by dynamics driven by the\textit{ standard (combinatorial) Laplacian} matrix: $A_{ij} - k_i \delta_{ij}$.  These include electrical networks \cite{huberman} and other systems with flow conservation \cite{maas}, or linearisations of oscillator networks~\cite{syn1,syn2}, among many others. Such a continuous-time stochastic process is driven by the standard Laplacian dynamics:
\begin{equation}
\label{ctrw2}
\dot{p}_{i} = \sum_{j} \frac{A_{ij}}{\langle k \rangle} p_{j} - \frac{k_i}{\langle k \rangle} p_i.
\end{equation}
This process stems from the random walk~(\ref{discrete}) by assuming
independent homogeneous Poisson processes at each node of the graph, but
now with \textit{ non-identical} rates that are proportional to the strength.
Hence, in contrast to process~(\ref{ctrw}), 
the probability to leave a node is proportional to its strength~\cite{random2} and its stationary solution
is uniform, $p_i^*=1/N$. The stability of a partition  based on the dynamics of process~(\ref{ctrw2}) is now given by
\begin{equation}
\label{Rt2}
R_{\rm CL}(t) =  \sum_C \sum_{i,j \in C} \biggl[ \left( e^{t (A-K)/\langle k \rangle} \right)_{ij} {1\over N} - {1\over N^2} \biggr],
\end{equation}
where $K_{ij}=k_j \delta_{ij}$. Again, keeping linear terms in $t$, one
obtains an approximate stability valid in the limit $t\rightarrow0$:
\begin{equation}
R_{\rm CL}(t) \approx (1-t) R_{\rm CL}(0) + t   Q_{\rm unif} \equiv Q_{\rm CL}(t),
\end{equation}
which is now directly related to the Erd\"os-Renyi (Bernoulli) version of modularity. 
Interestingly, $Q_{\rm CL}(t)$ is exactly equivalent both to the multi-resolution Hamiltonian in~\cite{reichardt0} (with
$t=1/\gamma$, $\gamma$ being the resolution parameter in that
method) and to the approach of Arenas et al.~\cite{stability1} based on the optimisation of $ Q_{\rm unif}$
for a modified network with $r$ self-loops added (with $t=\langle k \rangle/(\langle k \rangle + r)$) \footnote{In their original work, Arenas et al. focus on $Q_{\rm conf}$, but their approach can be generalized to other null models.}. 

The Laplacian processes (\ref{ctrw}) and (\ref{ctrw2}) are different except
for regular graphs with homogeneous strength $k_i=\langle k \rangle$, leading
to different stabilities $R_{\rm NL}(t)$ and $R_{\rm CL}(t)$ in general. This underscores the fact 
that there is no unique way to define the best partition of a network and that different quality functions will be more
or less appropriate depending on the nature of the network
and the dynamical processes underlying the system. For instance, the standard Laplacian~(\ref{ctrw2}) describes the (linearised) approach toward synchronisation of the Kuramoto model with identical intrinsic frequencies. Tracking the
transients of linearised Kuramoto dynamics has been used to uncover hierarchies in networks~\cite{syn1}. Interestingly, it has been observed~\cite{arenas2} that the partitions so uncovered optimise $Q_{\rm conf}$ only for homogeneous networks; that is, when $Q_{\rm conf}=Q_{\rm unif}$. This is expected from
our analysis since $Q_{\rm unif}$ is actually optimised by the dynamics. 

Our work indicates that different quality functions for partitions
can be associated with linear, stable, conservative dynamics taking place
on graphs through the consideration of the corresponding continuous-time
stochastic processes. The stabilities $R_{\rm NL}(t)$ and
$R_{\rm CL}(t)$ introduced above
are just two archetypical examples (see Supplementary Information for other examples) linked to standard forms of Laplacian dynamics concomitant with numerous physical and stochastic processes~\cite{chung}.
Based on the nature of the network
to be examined, this dynamical interpretation can aid in the definition of the most appropriate quality function to unfold the intrinsic sub-structure
of the network at different scales.

\newpage
\section*{Methods}

\noindent{\bf Optimization}

Several community detection methods use modularity as an objective function to be optimised in order to find the best partition of a network~\cite{N}.
Although it has been shown that optimization of modularity is NP-complete~\cite{NP}, several heuristic algorithms have been proposed to provide good approximations~\cite{fort}. We will now show that it is always possible to rewrite the stability of a graph as the modularity of another symmetric graph. This observation has important implications, as it is therefore possible to use any algorithm, e.g. spectral or greedy, for the optimisation of modularity in order to optimise stability.
In the case of a directed network, for example, where we have shown that stability $R(1)$ fundamentally differs from modularity, one can rewrite (\ref{R1}) as 
$
R(1) =  \sum_{C \in \mathcal P} \sum_{i,j \in C} \biggl[Y_{ij} - \pi_i \pi_j \biggr] 
$,
where the matrix $Y= \frac{X + X^T}{2}$ is manifestly symmetric, and where $X_{ij} =\frac{A_{ij}}{k_j^{out}} \pi_j$ is the flow of probability from $j$ to $i$ at stationarity. Because the inflow and outflow of probability at a node are equal at stationarity, network $X$ is Eulerian, i.e.,  $\sum_{j} X_{ij}= \pi_i = \sum_{j} X_{ji}$, and the strength of a node $i$ in network $Y$ is $\pi_i$. This implies that $R(1)$ is equal to the modularity of $Y$.

This observation is not only valid at time $t=1$ but can be generalized to any value of $t$. In the case of the continuous time process (\ref{ctrw}), for instance, $R_{\rm NL}(t)$ is equal to the modularity of a time-dependent weighted graph with adjacency matrix $X_{ij}(t)=\left(e^{t (B-I)} \right)_{ij} k_j$, which is symmetric due to detailed balance at equilibrium. This interpretation provides the following intuition: by construction, the matrix $X(t)$ explores larger and larger parts of the network while giving less and less weight to paths of length 1 (the links of the original matrix $A$). It is therefore expected that larger communities are found by optimising modularity of the weighted matrix $X(t)$ as time is increased. The stability of a graph can therefore be optimized by first evaluating $X$ and then by optimizing its modularity. In practice, the exponential of the transition matrix is evaluated via Pad{\'e} approximation \footnote{All the codes are available on {\it http://www.lambiotte.be.}}, and the optimization of modularity is performed by using a greedy algorithm~\cite{optim_method}. Let us insist on the fact that partitions at different values of $t$ are found independently by following the described procedure. In the case of very large networks, for which such an optimisation or the evaluation of $X$ is too onerous, it is instead possible to optimize the linearized stability $Q_{\mathcal M}(t)$, which can now be optimised at the same cost as modularity, i.e., $O(N)$ for the fastest algorithms on sparse graphs. We have implemented two such algorithms based on a deterministic greedy heuristic~\cite{Blondel} and stochastic simulated annealing~\cite{guimera}. 

\noindent{\bf Analyzing the sequence of partitions}
The optimization of stability over a period of time leads to a sequence of partitions that are optimal at different time scales.  The extraction of these partitions is a first step in order to uncover the multi-scale modular structure of the network, but it still requires at least two non-trivial steps.

On the one hand, one needs a way to select the most relevant scales of description, which is a well-known problem of multi-resolution methods. The significance of a particular scale is usually associated to a certain notion of the robustness of the optimal partition. Here, robustness indicates that  a small modification of the optimization algorithm\cite{ronhovde}, of the network \cite{stability1,karrer} or of the quality function \cite{reichardt} does not alter this partition. In this paper, we will follow the latter approach by defining the robustness of an optimal partition as its persistence over long periods of times. The persistence of a partition can be determined by looking for plateaux in the time evolution of summary statistics such as the number of communities. However, it is preferable and less ambiguous to actually compare the partitions and look for time intervals over which partitions are very similar \cite{mason}. A popular way to compare two partitions $\mathcal{P}_1$ and $\mathcal{P}_2$ is the normalized variation of information \cite{meila} 
\begin{equation}
\hat{V}(\mathcal{P}_1,\mathcal{P}_2)\equiv \frac{H(\mathcal{P}_1|\mathcal{P}_2)+H(\mathcal{P}_2|\mathcal{P}_1)}{log N},
\end{equation}
where $H(\mathcal{P}_1|\mathcal{P}_2)$ is the conditional entropy of the partition $\mathcal{P}_1$ given $\mathcal{P}_2$, namely the additional information needed to describe $\mathcal{P}_1$ once $\mathcal{P}_2$ is known. The conditional entropy is defined in a standard way for the joint distribution $P(C_1,C_2)$ that a node belongs to a community $C_1$ of $\mathcal{P}_1$ and to a community $C_2$ of $\mathcal{P}_2$. The normalized variation of information, which has been shown to be a true metric on the space of partitions, belongs to the interval $[0,1]$ and vanishes only when the two partitions are identical.  Relevant time scales are identified as block diagonal regions where $\hat{V}(\mathcal{P}_t,\mathcal{P}_{t^{'}})$ is significantly small.

On the other hand, one needs to check whether or not the sequence of partition is compatible with a hierarchical organisation. This problem requires the introduction of a quantity that measures whether the communities at some time $t^{'}$ are nested into the communities at a subsequent time $t>t^{'}$. A well-known information theoretic measure is particularly adapted for such a purpose, namely the normalized conditional entropy 
\begin{equation}
\hat{H}(\mathcal{P}_t|\mathcal{P}_{t^{'}})\equiv \frac{H(\mathcal{P}_t|\mathcal{P}_{t^{'}})}{log N},
\end{equation}
which again belongs to the interval $[0,1]$, but is now an asymmetric quantity that vanishes only if each community of $\mathcal{P}_t$ is the union of  communities of $\mathcal{P}_{t^{'}}$. The combined knowledge of  $\hat{V}$ and $\hat{H}$ therefore allows us to uncover the significant partitions of the system and to verify if those partitions are organized in a hierarchical manner.

\newpage

\section*{Supplementary Information}
\subsection*{Modularity for weighted or directed networks}

By definition, modularity tells us when there are more edges within communities than we would expect on the basis of chance. In general, this definition can be summarized by the formula

\begin{align}
Q &= \mbox{(fraction of edges within communities)} \nonumber\\
  &\qquad{} - \mbox{(expected fraction of such edges)}.
\label{eq:modDeff}
\end{align}
which leads to 
\begin{equation}
\label{modularity}
Q = {1\over2m} \sum_{C \in \mathcal{P}} \sum_{i,j \in C} \biggl[ A_{ij} - P_{ij} \biggr]
\quad \quad
\begin{cases}
\text{if $P_{ij}=\langle k\rangle^2/2$,} & \text{then $Q \equiv   Q_{\rm unif}$} \\
\text{if $P_{ij}= k_i k_j/2m$,} &  \text{ then $Q \equiv  Q_{\rm conf}$.}
\end{cases}
\end{equation}
 in the case of undirected and unweighted networks. The null model $P_{ij}= k_i k_j/2m$ is usually preferred because it captures the degree heterogeneity of the network. For general networks, the main difficulty consists in choosing a null model consistent with the network under consideration. In the case of undirected and weighted networks, modularity is usually defined as
\begin{equation}
\label{modularityw}
Q = {1\over2m} \sum_{C \in \mathcal{P}} \sum_{i,j \in C} \biggl[ A_{ij} - \frac{k_i k_j}{2m} \biggr],
\end{equation}
where $A_{ij}$ is now the weight of a link between $i$ and $j$, and $k_i=\sum_j A_{ij}$ is the strength of node $i$. For directed networks, it has been proposed to modify the null modified in order to account for the directionality of the links, thereby leading to the expression \cite{leichl,Arenas}
\begin{equation}
\label{modularitydir}
Q = {1\over2m} \sum_{C \in \mathcal{P}} \sum_{i,j \in C} \biggl[ A_{ij} - \frac{k_i^{in} k_j^{out}}{2m} \biggr],
\end{equation}
where $k_i^{in}=\sum_j A_{ij}$ and $k_i^{out}=\sum_j A_{ji}$.

\subsection*{ Stability for directed networks.}

As we stressed in the main body of this article, stability $R_{\mathcal{M}}(t)$ and modularity $Q$ differ in a fundamental way in the case of directed networks, and are in principle optimized by different partitions. While $R_{\mathcal{M}}(t)$ favours partitions with long persistent flows of probability within modules, (\ref{modularitydir}) favours partitions with high densities of links and is blind to the flow actually taking place on these links. In order to illustrate the difference, let us focus on the example taken from \cite{rosvall}. Optimising the modularity of this toy network leads to a partition where heavily weighted links are concentrated inside communities, as expected. Optimising stability, instead, leads to a partition where flows are trapped within modules. It is also interesting to stress that the partition optimising  $R_{RW}(1)$ also optimises the map equation proposed by Rosvall and Bergstrom\cite{rosvall}. For an independent study of $R_{RW}(1)$, we refer to the recent work of Kim et al\cite{fKim}.

Our definition of stability relies on the condition that the dynamics is ergodic.
When the directed network is not ergodic, it is common to generalise the standard random walk (\ref{discrete}) by incorporating random teleportations. If the walker is located on a node with at least one outlink, it follows one of those outlinks with probability $1-\tau$. Otherwise, the random walker teleports with a uniform probability to a random node. The corresponding transition matrix from $j$ to $i$ is given by
 \begin{equation}
 \label{google}
(1-\tau) \frac{A_{ij}}{k_j^{out}} + \frac{1}{N} ((1-\tau) a_j + \tau),
\end{equation}
where $a_j$ is equal to 1 if $j$ is a dangling node. This scheme is known to make the dynamics ergodic and to ensure the existence of one single stationary solution $\pi_i$ that is an attractor of the dynamics.

\subsection*{  Resolution limit and size dependence.}
The resolution limit of modularity \cite{FB} is a well studied phenomenon that imposes
a limit on the size of the smallest community one can obtain by modularity
optimisation. It is a fundamental issue emerging from the definition of modularity.
In particular, it originates from the factor $1/2m$ in the null model $P_{ij}$ which implies that modularity depends on the size of the network and not only on its local properties. It has been shown that the tuneable Hamiltonian of Reichardt and Bornholdt \cite{reichardt}, and hence our $Q_{NL}(t)$, has a resolution limit for any given value of the resolution parameter $\gamma$~\cite{jari}. 
However, the approximate quality measure $Q_{NL}(t)$ has a remarkable property
that implies that a change in the size of the system simply translates into a change in the time at which a partition is optimal. Let $\mathcal{P}_1$ be the optimal partition of a network $\mathcal{N}_1$
for some value of $t$, i.e., $\mathcal{P}_1$ optimises the quality function
\begin{equation}
\label{n1}
Q_{NL}(t) = (1-t) +  \sum_{C \in \mathcal{P}} \sum_{i,j \in C}  \biggl[  \frac{A_{ij}}{2m_1} t - \frac{k_i k_j}{(2m_1)^2} \biggr],
\end{equation}
where $m_1$ is the total number of links in $\mathcal{N}_1$.
 Let us now consider a new network $\mathcal{N}_{tot}$ made of the union of $\mathcal{N}_1$ and of another network $\mathcal{N}_2$ such that there are no links between $\mathcal{N}_1$ and $\mathcal{N}_2$. Let us now show that $\mathcal{P}_1$ still belongs to the optimal partition of $\mathcal{N}_{tot}$, but for another value of time. To do so, one should first note that the optimal partition of $Q(t^{*})$ is made of connected communities, so that the optimal partition of $\mathcal{N}_{tot}$ is a union of partitions $\mathcal{P}_1^{*}$ and $\mathcal{P}_2^{*}$ of $\mathcal{N}_{1}$ and $\mathcal{N}_{2}$ respectively, i.e., all the nodes of a community either belong to $\mathcal{N}_1$ or to $\mathcal{N}_2$. This implies that $\mathcal{P}_1^{*}$ optimises the quantity
\begin{equation}
\label{total}
 \sum_{C \in \mathcal{P}} \sum_{i,j \in C \subseteq \mathcal{N}_1}  \biggl[  \frac{A_{ij}}{2m_{tot}} t^{*} - \frac{k_i k_j}{(2m_{tot})^2} \biggr],
\end{equation}
where $m_{tot}$ is now the total number of links in
$\mathcal{N}_{tot}$ and where $A_{ij}$ is the same as in (\ref{n1})
by construction. By comparing (\ref{n1}) and (\ref{total}), it is
now easy to show that $\mathcal{P}_1$ also maximises (\ref{total})
and therefore belongs to the optimal partition of
$\mathcal{N}_{tot}$ for a value of time $t^{*}=t\, (m_{1}/ m_{tot})$. 
Consequently, changing the size of the system only alters the
resolution of the method. A similar relation can be obtained for
$Q_{CL}(t)$.

\subsection*{  Partitions at long times.}  In order to reveal the
properties of $R_{NL}(t)$ in the limit $t \rightarrow \infty$, let us
consider the spectrum of the matrix $e^{(B-I) t}$. 
It is straightforward to show that its
eigenvalues are real and positive. The largest eigenvalue
$\lambda_1$ is equal to 1 and its associated eigenvector $v^{(1)}_i=k_i/2m$
corresponds to the stationary solution of the random process. All
the other eigenvalues are associated with relaxations toward
stationarity and have the form $\lambda_k=e^{-a_k t}$, with $a_k>0$
and therefore vanish when $t \rightarrow \infty$. If we denote by
$\lambda_k$ the $k^{\mathrm{th}}$ largest eigenvalue and by 
$v_i^{(k)}$ its corresponding eigenvector, one can then rewrite $R_{NL}(t)$ as
\begin{equation}
\label{vector}
R_{NL}(t) = \sum_C \sum_{i,j} u^{(C)}_{i} \biggl[ \left(e^{t (B-I)}\right)_{i,j}
- {k_i\over2m} \biggr] {k_{j}\over2m} u^{(C)}_{j},
\end{equation}
where the elements of the vectors $u^C_{i}$ are $1$ if $i\in C$ 
and $0$ otherwise. The vector ${k_{j}\over2m} u^C_{j}$ may be decomposed in the basis of eigenvectors $v^k_j$. The contribution of the first eigenvector
$v^1_{j}$ is shown to vanish for all times and for any partition. In
the limit $t\rightarrow \infty$, the main contribution 
comes from $v^2_{j}$ and $R_{NL}(t)$ is shown to be optimised 
by the two-way Fiedler-type partition, deduced from the sign of
entries of the second eigenvector of the normalized Laplacian~\cite{shimalik}.
The optimal partitions of $R_{CL}(t)$ when $t \rightarrow \infty$ are obtained in the same way except that they are now based on the second eigenvector of the combinatorial Laplacian, as originally proposed by 
Fiedler~\cite{fiedler}. Hence, at long times, $R_{NL}(t)$ and $R_{CL}(t)$ recover the two classical spectral algorithms based on the two Laplacians.
 
\subsection*{More general random walks}

The continuous-time random walks described in this paper are based on the same discrete-time process 
\begin{equation}
\label{discrete}
p_{i;n+1} = \sum_{j} \frac{A_{ij}}{k_j} p_{j;n},
\end{equation}
where a walker follows a links with a probability proportional to its weight, and only differ in the statistics of the waiting times between two jumps. \begin{equation}
\label{ctrw}
\dot{p}_{i} = \sum_{j} \frac{A_{ij}}{k_j} p_{j} - p_i
\end{equation}
and 
\begin{equation}
\label{ctrw2}
\dot{p}_{i} = \sum_{j} \frac{A_{ij}}{\langle k \rangle} p_{j} - \frac{k_i}{\langle k \rangle} p_i
\end{equation}
 are therefore two particular cases of the continuous time process
\begin{equation}
\label{ctrwgeneral}
\dot{p}_{i} = \sum_{j} \frac{A_{ij}}{k_j} \tau(k_j) p_{j} - \tau(k_i) p_i,
\end{equation}
where $\tau(k_i)$ is the rate at which random walkers leave a node of degree $k_i$ and whose stationary solution is $p_i^*=k_i/\tau(k_i)$.

\bigskip
\noindent {\bf Acknowledgments}

We thank T.\ Evans, S.\ Yaliraki, M.\ Draief, H.J.\ Jensen, V.\ Blondel and J.\ Saram\"aki for fruitful discussions and helpful comments. R.L. acknowledges support from the UK EPSRC. J.-C.D. is supported by a  FNRS fellowship, the Belgian Programme on Interuniversity Attraction Poles and an ARC of the French Community of Belgium.

Correspondence and requests for materials should be addressed to R.L. and
M.B.

\newpage
\section*{Figure Legends}

\noindent
{\bf Figure 1: Stability of the partition of a network.}
Given a network and a partition of the network (into two communities indicated by the red and green colors), one can evaluate the quality of the partition from the statistical properties of coarse-grained trajectories of a random
walker. A partition will be of high quality at a given time-scale if the
walker has high probability of remaining within the specified communities
within that time-scale (left). If the partition is of low quality (right),
the walker will switch at a high rate between communities.  The stability of the partition $R(t)$, defined in~(\ref{Rt}), can be interpreted as the time auto-correlation of such a coarse-grained
signal, where a numerical value is assigned to each colour. This measure gives information about the average time spent by the walker inside the specified communities and establishes a time-dependent measure of the quality of a given partition.

\bigskip
\noindent
{\bf Figure 2: Sequence of optimal partitions of a hierarchical network as a function of time.}
We consider a regular hierarchical graph generated as follows~\cite{hier}. Start with a pair of nodes connected by a link
of weight $c<1$, duplicate them and add a
link of weight $c^2$ between all pairs of nodes in
different modules. This meta-module of four nodes is
duplicated and links of weight $c^3$ are added between nodes of
different meta-modules. A fully-connected, weighted network  of $2^K$ nodes is
obtained by iterating this step $K$ times.
In the right upper corner, we represent such a network with $2^4=16$ nodes and edges shaded according to their strength ($c=1/4$).
By symmetry, the natural partitions are into $16$ single nodes, $8$ pairs of nodes (colours), $4$ groups of $4$ nodes (shapes) and $2$ groups of $8$ nodes (upper and lower hemispheres). The figure also shows the stability $R(t)$ of the natural partitions of
this graph. As $t$ grows, the sequence of optimal partitions goes from 16 communities to 8 to 4 to 2. 

\bigskip
\noindent
{\bf Figure 3: Sequences of optimal partitions of the karate network.} 
We optimize the stabilities $R_{NL}(t)$ and $R_{CL}(t)$ of the karate network ~\cite{karate}, which is a small social network made of 34 nodes.
(a) Normalised variation of information $\hat{V}(\mathcal{P}_t,\mathcal{P}_{t^{'}})$ between the optimal partitions of $R_{NL}$ at different times $t$ and $t^{'}$. We also plot the number of communities of the corresponding partitions. It is interesting to note that a constant number of communities does not imply that the partitions are equivalent, and that variations in the number of communities may have a marginal effect on the distance between the partitions. The most persistent partitions are highlighted by dashed lines. (b) The small values of the normalized conditional entropy $\hat{H}(\mathcal{P}_t,\mathcal{P}_{t^{'}})$ indicate that partitions at different times are nested in each other. (d) Normalised variation of information $\hat{V}(\mathcal{P}_t,\mathcal{P}_{t^{'}})$ for $R_{CL}$. A comparison of (a) and (d) shows that  $R_{NL}(t)$ and $R_{CL}(t)$ are optimized by distinct sequences of persistent partitions. This is confirmed by looking in (c) at $\hat{V}(\mathcal{P}_t,\mathcal{P}_{t^{'}})$ where $\mathcal{P}_t$ now denote optimal partitions of $R_{NL}(t)$ and $\mathcal{P}_{t^{'}}$ optimal partitions of $R_{CL}(t^{'})$. The differences between both measures emanate from the underlying dynamical processes with distinct stationary distributions that link $R_{NL}(t)$ to the configuration null model and $R_{CL}(t)$ to the uniform null model.

\bigskip
\noindent
{\bf Figure 4: Sequences of optimal partitions of a benchmark network.} 
The benchmark network is made of $640$ nodes with $3$ hierarchical levels (modules of $10$, $40$ and $160$ nodes respectively) \cite{sales}. The density of links within the modules at different levels is tuned by a single parameter $\rho$, $\rho=1.5$ in this example. 
In (a) and (b), we plot the normalised variation of information $\hat{V}(\mathcal{P}_t,\mathcal{P}_{t^{'}})$ and normalized conditional entropy $\hat{H}(\mathcal{P}_t,\mathcal{P}_{t^{'}})$ for the optimal partitions of $R_{NL}(t)$ and $R_{NL}(t^{'})$. In (a), we also plot the number of communities of the corresponding partitions and, in dashed lines, the number of communities in the natural partitions.
 The optimal partitions are all evaluated for the same realization of the random graph.
Because the network is homogeneous (and therefore almost regular), we do not focus on $R_{CL}(t)$.

\bigskip
{\bf Figure SI: Modularity vs stability.}
In this toy network proposed by Rosvall and Bergstrom\cite{rosvall}, the weight of the bold links is twice the weight of the normal links. The partition on the left is shown to optimise $R_{RW}(1)$. The partition on the right instead optimises modularity.

\newpage
\begin{figure}[!ht]
\begin{center}
\vskip .7cm
\includegraphics[width=12cm]{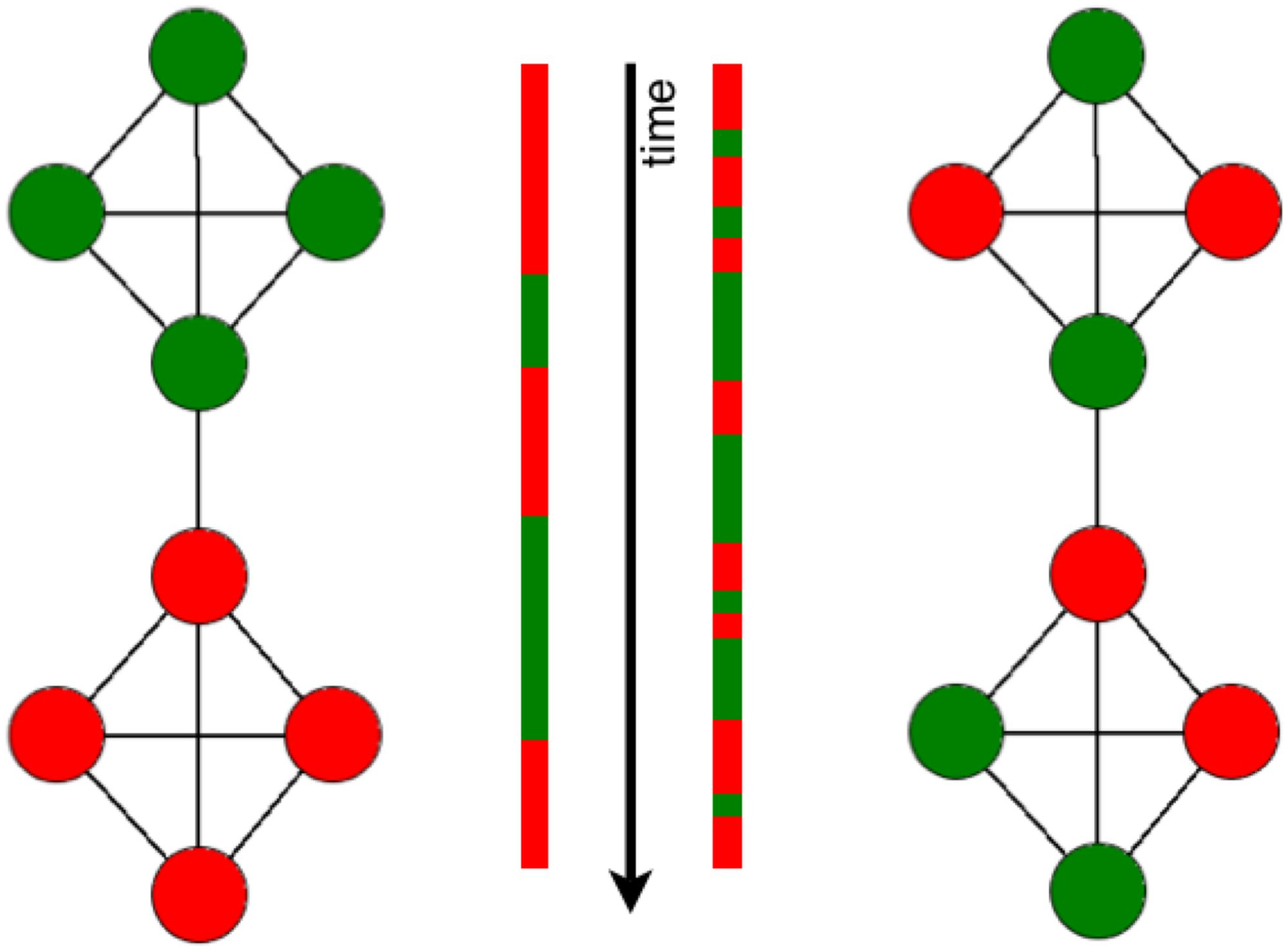}
\vspace{.3cm}

{\bf Figure 1.}
\end{center}
\label{fig1}
\end{figure}

\newpage
\begin{figure}[!ht]
\begin{center}
\vskip 2cm
\includegraphics[height=8cm]{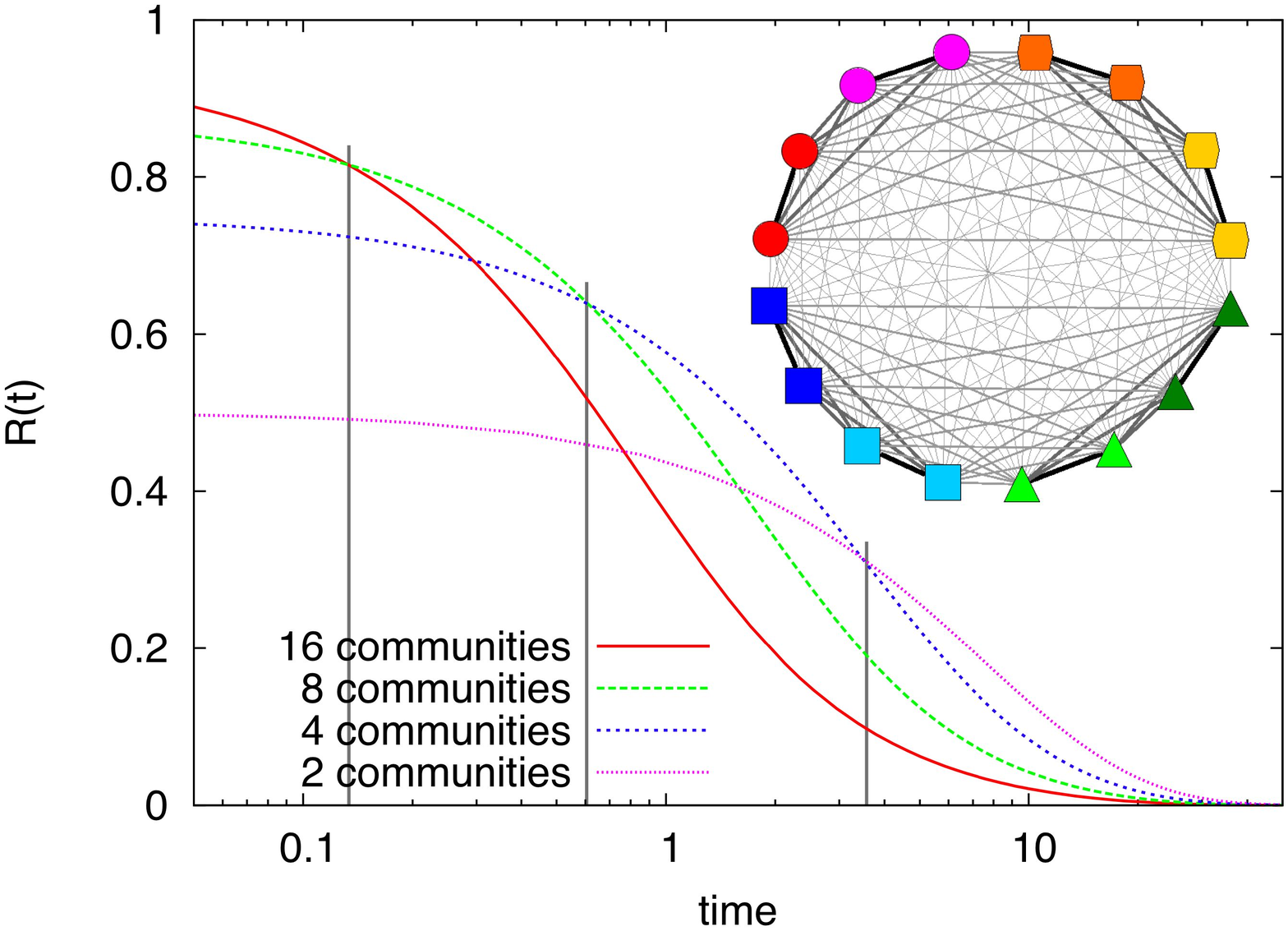}
\vskip .1cm
\renewcommand{\baselinestretch}{1.0}
\vspace{.3cm}

{\bf Figure 2.}
\end{center}
\label{fig2}
\end{figure}

\newpage
\begin{figure}[!ht]
\begin{center}
\vskip 2cm
\includegraphics[height=12cm]{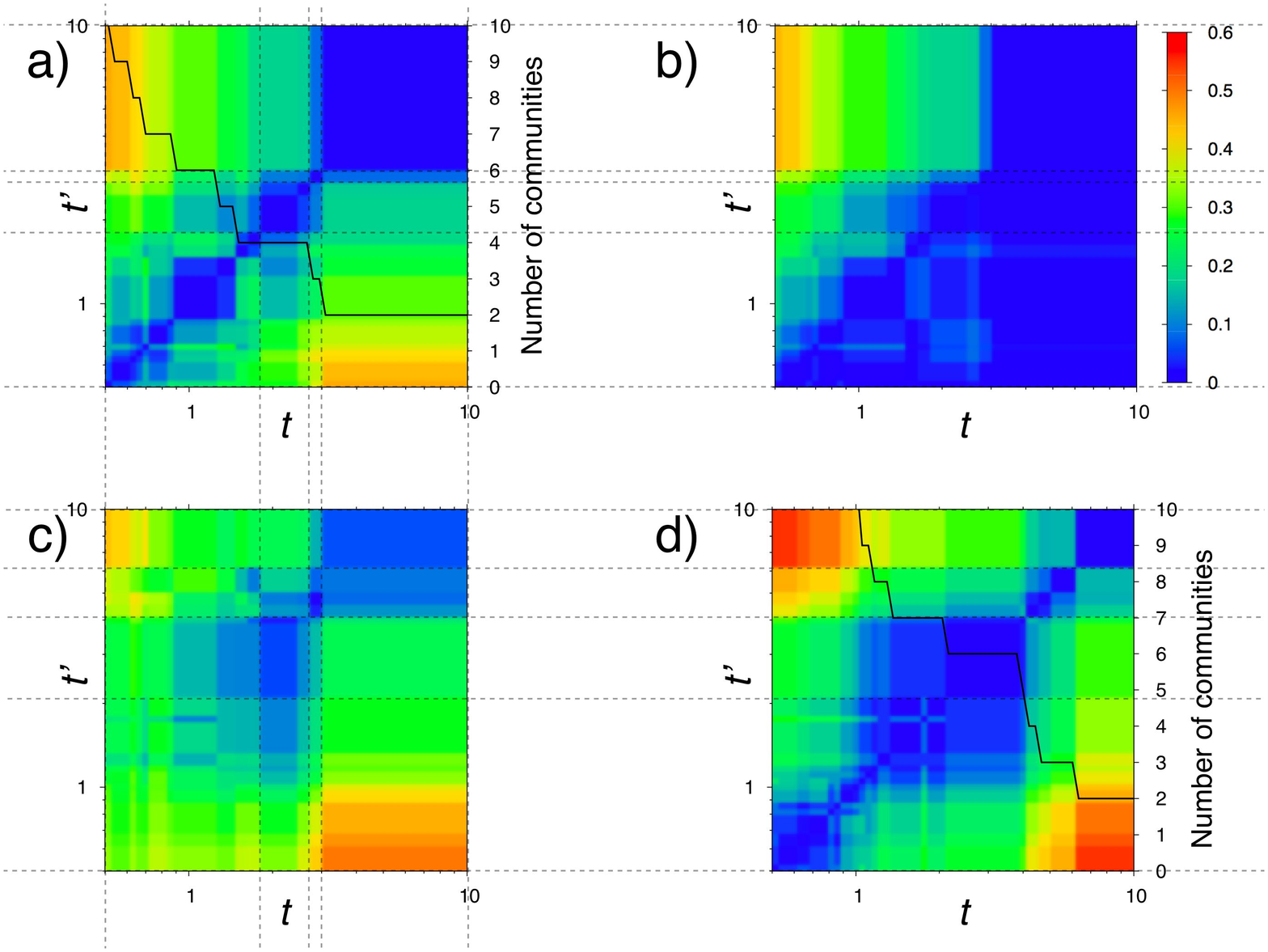}
\vskip .1cm
\renewcommand{\baselinestretch}{1.0}
\vspace{.3cm}

{\bf Figure 3.}
\end{center}
\label{fig2}
\end{figure}

\newpage
\begin{figure}[!ht]
\begin{center}
\includegraphics[width=8cm]{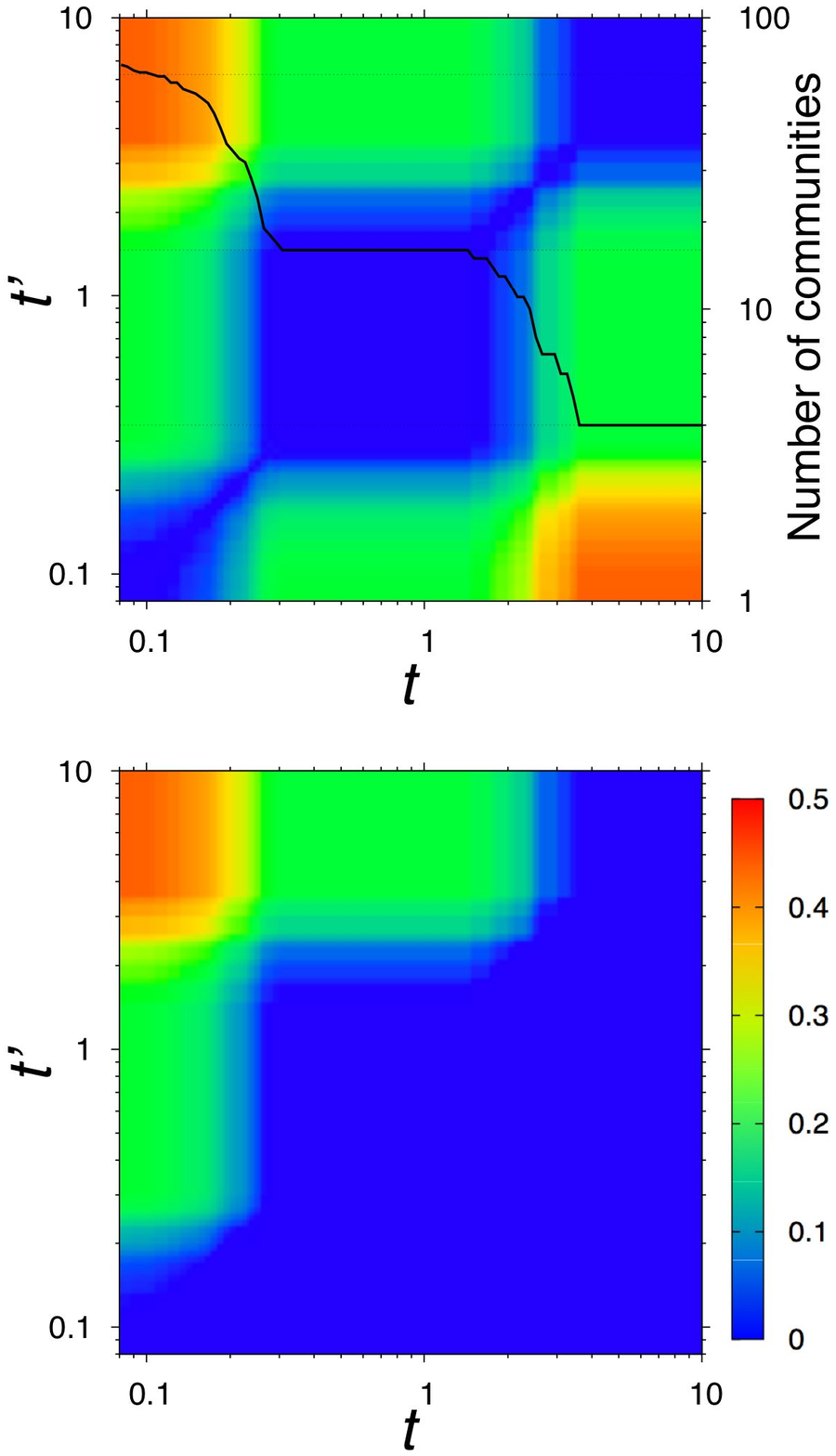}
\renewcommand{\baselinestretch}{1.0}
\vspace{.3cm}

{\bf Figure 4.}
\end{center}
\label{fig3}
\end{figure}

\newpage
\begin{figure}[!ht]
\begin{center}
\vskip .7cm
\includegraphics[width=12cm]{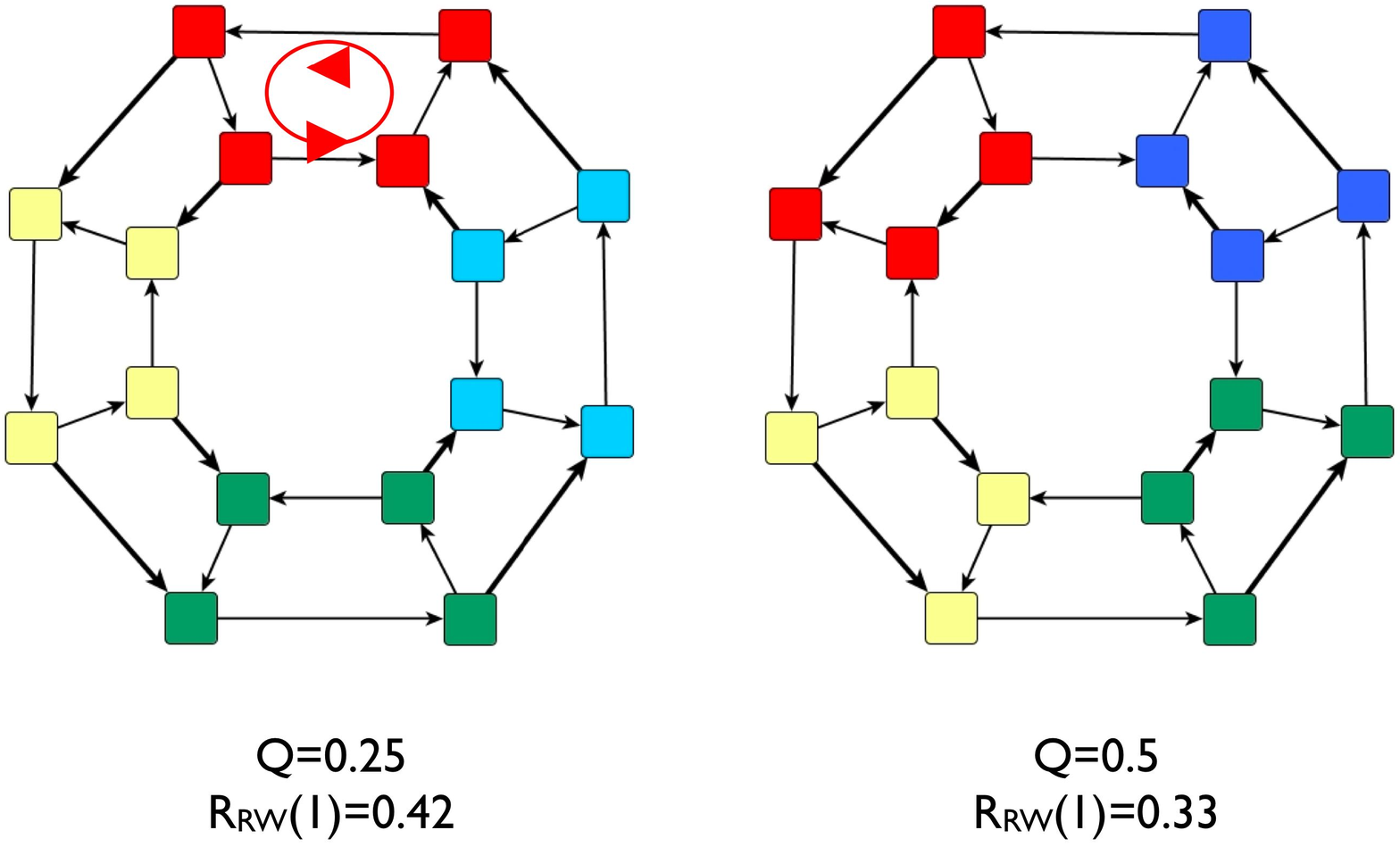}
\vspace{.3cm}

{\bf Figure SI.}
\end{center}
\label{figSI}
\end{figure}

\end{document}